# A Simple and Precise Way to Determine Electrical Resistivity of Isotropic Conductors: Simplifying the Four-Probe Method

L. M. S. Alves[1,2], F. S. Oliveira[1,3], E. C. Romão[1], M. S. da Luz[4], and C. A. M. dos Santos[1,a]

[1]*Escola de Engenharia de Lorena, Universidade de São Paulo, Lorena - SP, 12602-810, Brazil*
[2]*Instituto Federal de Educação, Ciência e Tecnologia Catarinense, Araquari - SC, 89245-000, Brazil*
[3]*Instituto de Física "Gleb Wataghin", Universidade Estadual de Campinas, Campinas - SP, 13083-970, Brazil*
[4]*Instituto de Ciências Tecnológicas e Exatas, Universidade Federal do Triângulo Mineiro, Uberaba - MG, 38025-180, Brazil*

[a]Author to whom correspondence should be addressed: cams-eel@usp.br.

COMSOL Multiphysics software is used to describe the behavior of the electrical resistivity of several samples with rectangular shape typically used in the Montgomery method. The simulation data obtained using four isotropic conductors allowed us to understand in detail the behavior of the electric potential and electric field of the samples. The results provide an analytical method which can substitute the four-probe method with much more simplicity and precision.

*Key words:* Electrical resistivity; Montgomery method; Four-probe method; COMSOL.

## I. INTRODUCTION

The electrical resistivity ($\rho$) of a given material is one of the most important physical properties and its correct determination has been the object of intense study during the last century.[1-5] This property has been used to classify materials into metals, semiconductors, insulators, and superconductors, which are the basis for electronic devices, such as resistors, capacitors, diodes, transistors, and many others.[6,7]

Several methods have been described for measuring the electrical resistivity.[8-19] One of them was first described in 1915 by Frank Wenner,[8] who was trying to measure the electrical resistivity of the Earth with a method based on in-line four-probes. Basically, the voltage drop V is measured between the two internal contacts, while an electrical current I is injected through the two external terminals. Later on, this method was used to measure the electrical resistivity of a semiconductor wafer, being established as a reference procedure of the American Society for Testing and for Materials Standards.[20] Nowadays, this method has been widely used to measure the electrical resistivity of conducting materials and it is well known as the conventional four probe method (4P method).[1]

In fact, 4P method is based upon the Ohm's law.[1] the four electrodes are usually arranged along to one of the faces of a rectangular sample, as shown in Figure 1, which must strictly respect some geometrical aspects.

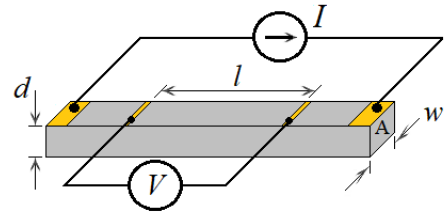

**FIG. 1.** Schematic view of the electrical contacts in the 4P method for a sample with rectangular shape, in which: i) the sample geometry must respect $\sqrt{A} \ll l$; ii) the voltage contacts should be aligned perpendicular to the distance l; and iii) the electrical current terminals must be large enough to have small contact resistances to avoid Joule self-heating.

Applying an electric field ($E$) to a conductor, an electric current ($I$) immediately appears and the electrical resistivity can be calculated from the ratio of the electric field and the current density ($J \equiv I/A$).[21] In fact, $\rho$ is not directly determined in the experiments, but calculated after measuring the electrical resistance ($R$), which is obtained from the ratio of the measured voltage drop and the applied current.

The electrical resistivity is calculated by

$$\rho = R\frac{A}{l} \qquad (1)$$

where $A = wd$ is the area of the specimen and $l$ is the distance between the two points where the voltage drop is measured.

In order to have precise determination of the electrical resistivity, the geometry of the sample and the electrical contacts must be well established. To measure the voltage drop, the electric field must be approximately constant, which is obtained cutting the sample with aspect ratio that respects $\sqrt{A} \ll l$. The voltage contacts should be point or line-like (see Figure 1 again), which must be perpendicular to the electrical field lines. Furthermore, both voltage and electrical current terminals must have very small contact resistance to avoid thermopower and Joule self-heating effects.

Although widely used, the 4P method has some limitations due these experimental requirements. Samples of interest to be measured are usually in the millimeter size range and many times are not bars. In addition, electrical contacts cannot be prepared as pointed out above. These typical limitations always provide electrical resistivity measurements which have errors as large as 10 to 50%.[12, 13]

In this work the problem of measuring the electrical resistivity of isotropic samples of finite dimensions using the four-probe techniques has been revisited. Our previous experience measuring transport properties of isotropic and anisotropic samples lead us to modified the Montgomery and the van der Pauw methods.[17, 18] Those works allowed to find ways to calculate the electrical resistivity of both isotropic and anisotropic samples analytically. However, in order to understand these methods deeper, we intend to apply them to several sample geometries of different conducting materials. This is not a simple experimental task. We have noticed that this could be done by using numerical simulations. This work reports our first simulation results obtained from COMSOL Multiphysics software for rectangular thin samples using the Montgomery method. The simulations provide a method which is simpler and more precise than the 4P method.

## II. METHODOLOGY

COMSOL Multiphysics software was used to carry out all the simulations shown in this work.[22] This commercial software is much used for numerical projects in several knowledge areas, such as electric field, heat and mass flow, fatigue, and environmental studies, among others.[23-28]

We have solved the problem of two dimensional (2D) stationary electrical currents in conductive media, considering the equation of continuity in a stationary coordinate system, based on the Ohm's law, which states that

$$J = \sigma E + J_e, \quad (2)$$

where $\sigma$ is the electrical conductivity (in S/m), $E$ is the electric field (in V/m), and $J_e$ is an externally generated current density (in A/m$^2$), which is taken zero in the simulations. In such a case, the static form of the equation of continuity is given by

$$\nabla \cdot J = -\nabla \cdot (\sigma \nabla V) = 0. \quad (3)$$

The boundary conditions adopted a thin rectangular domain with dimensions $-L_1/2 \leq x \leq L_1/2$, $0 \leq y \leq L_2$, and thickness $L_3 = 1$ m, which respects $L_3 < \sqrt{L_1 L_2}$. Thus, the boundary conditions were determined in the following way:

$$V_A = V(-L_1/2, L_2) = -100 \text{ V} \quad (4a)$$

$$V_B = V(L_1/2, L_2) = 100 \text{ V} \quad (4b)$$

and for any other point in the boundary, the simulations took

$$n \cdot J = 0, \quad (4c)$$

where $n$ is normal vector at the surface boundary.

In the COMSOL software the distributions of electric field lines as well as equipotential lines were easily simulated and the values of the $V$, $E$, $I$, and $J$ as a function of the positions were recorded. Some schematic 2D view of the equipotential and electric field lines were also obtained. For example, Figure 2 displays these lines for a typical isotropic conductor with rectangular thin geometry

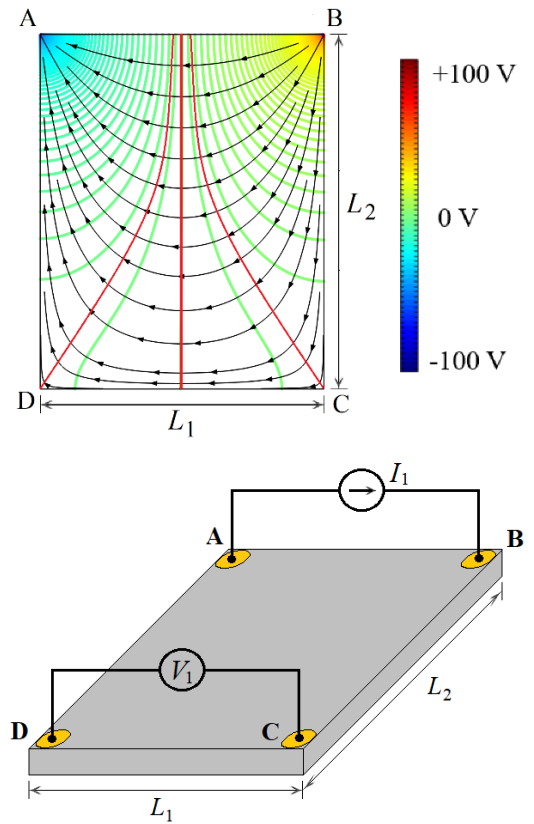

**FIG. 2.** (Upper) Typical schematic 2D view of the equipotential and electric field lines for a rectangular isotropic thin sample of dimensions $L_1$ and $L_2$ obtained using COMSOL multiphysics software. Input electric potentials are applied at A and B poles and output voltage is measured at C and D corners. The scale with color represents the voltage change over the entire simulated area. The highlighted vertical red line represents the zero equipotential line (at $x = 0$) and divided the rectangular sample into two symmetric parts. The other two red lines touching the points C = $(L_1/2, 0)$ and D = $(-L_1/2, 0)$ are equipotential lines which provide $V_1 = V_C - V_D$, which is crucial for the Montgomery method. (Lower) It is shown the typical sample geometry used in the Montgomery method.

To verify the distribution of the equipotential and electric field lines across the samples, the electric potentials of $-100$ V and $+100$ V (Equations 4a and b) were used in the A and B poles, as shown in Figure 2. They simulated the application of the electrical current in the both poles as done in the Montgomery experiments (see the lower Figure 2), whose value is adjusted by the applied voltage ($V = V_B - V_A = 200$ V) and the electrical resistance of the samples, which depends on the conducting material and the geometrical parameters $L_1$, $L_2$, and $L_3$.

## III. RESULTS AND DISCUSSION

Figure 3 displays the electric potential ($V$) and electric field ($E$) for four rectangular thin samples with different aspect ratios $L_1$ and $L_2$. The values of the potential and electric field were collected by performing a sweep around the A-D-C-B path shown in Figure 2a.

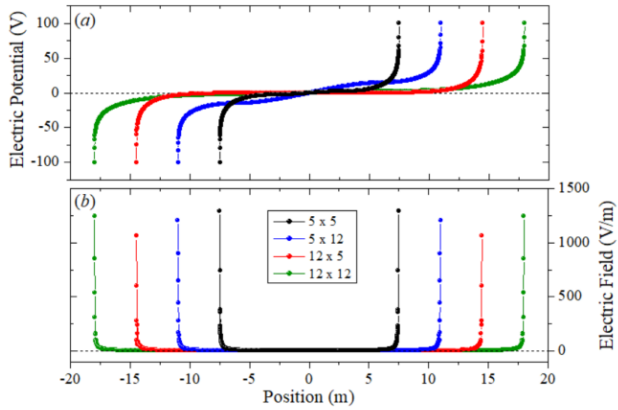

**FIG. 3.** Electric potential ($a$) and electric field ($b$) as a function of position along the A-D-C-B path for four different samples. The origin was defined at the midpoint of the D-C line.

Near the negative and positive poles (A and B corners) the values of electric fields and potentials vary strongly as a function of the position. On the other hand, in the opposite side of the samples (D-C path), both $V$ and $E$ are orders of magnitude smaller. Furthermore, it is observed that the shape of the curve for the different sizes are very similar and can be easily collapsed (not shown).

The Figure 4 shows the electric field and electric potential in logarithm scales for the A-D-C-B path of the sample with $L_1 = 12$ m and $L_2 = 5$ m.

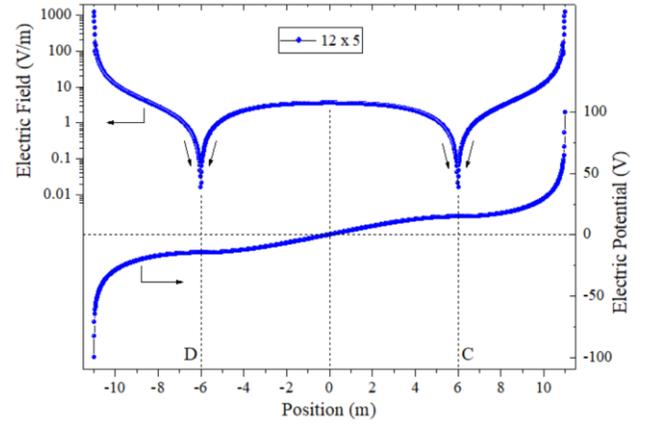

**FIG. 4.** Electric field (upper curve) and electric potential (lower curve) along A-D-C-B path for the sample with $L_1 = 12$ m and $L_2 = 5$ m.

Electric field (upper curve) is related to electric potential (lower curve) by $\vec{E} = -\nabla V$. Interesting is to note that the electric potential has saddle points at corners C and D, where the electric field vanishes (see arrows). At the middle point of the D-C path ($x = 0$), the electric potential is maximum and the electrical potential is zero, as expected due to the symmetric arguments. The same general behavior has been observed for all the simulated samples.

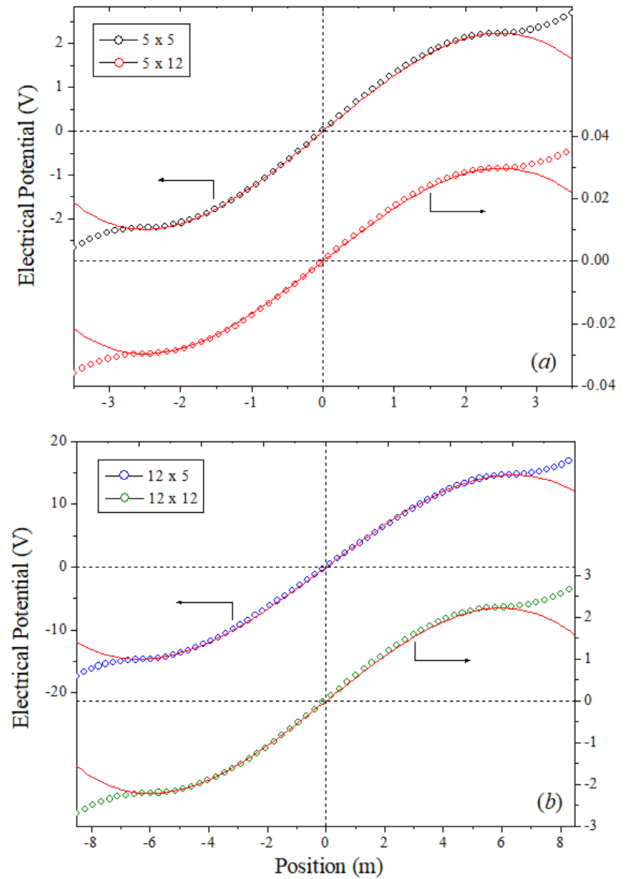

**FIG. 5.** Electric potential along the D-C path for the samples $L_1 \times L_2 = 5 \times 5$, $5 \times 12$, $12 \times 5$, and $12 \times 12$. Red solid lines are fits based on Equation (5).

To have a deep insight about the electric behavior in the D-C path, in Figure 5 is plotted the electric potential for the four samples.

Considering the zero electric potential at $x = 0$ and the two saddle points at $x = \pm L_1/2$, the behavior of the $V(x)$ must be described by a third order polynomial, whose boundary conditions impose that

$$V(x) = \frac{V_1}{2}\left(3\frac{x}{L_1} - 4\frac{x^3}{L_1^3}\right) \qquad (5)$$

Figure 5 also displays the fits based on Equation (5) (red solid lines). The agreement is excellent and allows one to determine the $V_1 = V_C - V_D$ value for each sample with high precision either by using the polynomial fit or from the saddle points.

In Figure 6 is shown a set of electric potential curves as a function of the position for samples with a fixed D-C distance ($L_1 = 5$ m) and different $L_2$.

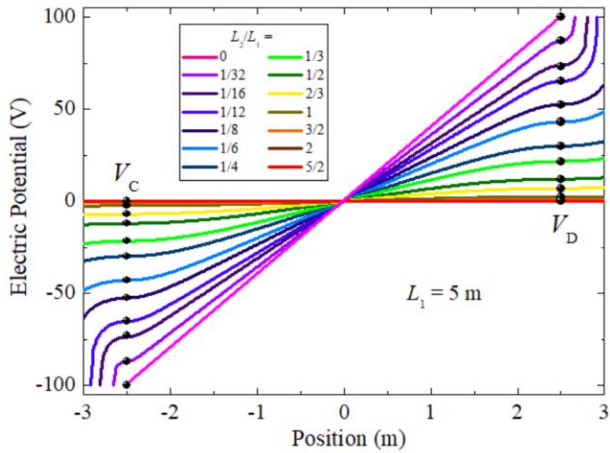

**FIG. 6.** Electric potential curves as a function of the position for samples with a fixed $D$-$C$ distance ($L_1 = 5$ m) and different $L_2$.

It is easy to notice that the saddle points are in the same position for all the samples as indicated by the small black solid circles at $x = \pm 2.5$ m. Similar behaviors have been observed for samples with different $L_1$ values, which demonstrate this is a general behavior of the rectangular samples in the Montgomery method.[14] As previously mentioned, this would be difficult to be performed experimentally. Thanks to COMSOL Multiphysics software which allowed us to do this type of simulation for many samples very fast.

As reported in previous paper,[14, 17] there is a simple relationship between the thin rectangular sample and its electrical resistance. This is given in the modified Montgomery method (see Equation (10) of the reference 17) by

$$\rho = (\pi/8)\, L_3\, R_1\, \sinh(\pi L_2/L_1), \qquad (6)$$

where $R_1 = V_1/I_1$ is the sample resistance in the Montgomery method and $L_3$ is the samples thickness. Interesting is the square sample ($L_1 = L_2$), in which sheet resistance is given by the value $\rho/L_3 = 4.535\, R_1$.[18]

In order to check the validity of the Equation (6) for the simulation data obtained in this work, in Figure 7(a) is plotted the $V_1$ values obtained as described in Figure 5 as a function of the $1/\sinh(\pi L_2/L_1)$ in the limit $L_2/L_1 \geq 0.5$ for four different metallic materials (Cu, Al, Pt, and Hg). For this particular case, a constant electrical current of 1 A was used in the simulations of the thin samples ($L_3 = 1$ m).

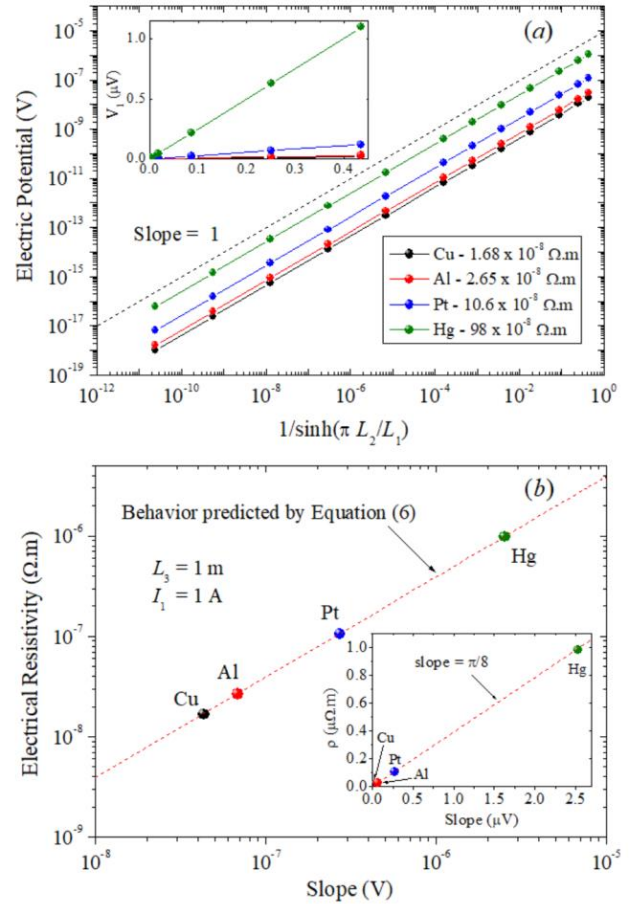

**FIG. 7.** (a) $V_1$ values obtained as described in the Figure 5 as a function of the $1/\sinh(\pi L_2/L_1)$ for the limit $L_2/L_1 \geq 0.5$, using simulations with a constant electrical current of 1 A. In (b) is plotted the electrical resistivity of the four metals as a function of the slope obtained from the inset of the Figure (a). The red dot line represents the behavior predicted by Equation (6).

The linear behavior observed in both log-log (main panel) and linear (inset) scales in the Figure 7(a) for all the metals unambiguously demonstrate that the Equation (6) describes the simulated data pretty well. Furthermore, by plotting the electrical resistivity as a function of the slope obtained from inset of the Figure 7(a), as displayed in Figure 7(b), shows an excellent agreement with pre-factor ($\pi/8$) predicted in the Equation (6).

Finally, we have noticed that the results shown in this work allow us to suggest a simple and precise method for determining the electrical resistivity of isotropic conductors, especially when sample is like a thin rectangular block ($L_1$ is comparable to $L_2$ and $L_3 \ll \sqrt{L_1 L_2}$). In such a case, the following advantages have been observed with regard to the conventional 4P method: *i)* contacts can be placed at the corners of the samples becoming easier to handle with and avoiding short circuits; and *ii)* precision to determine electrical resistivity can be improved since the samples can be bigger and contact smaller than in the 4P method, as well as avoids misorientation in the contacts.

Just to check the method proposed here, we have revisited the experimental data reported previously for copper plates and aluminum foils (see reference 17) to compare with the sheet resistance ($\rho/L_3$) predicted by Equation (6). The agreement is excellent demonstrating that is easy to use Equation (6) for determining the electrical resistivity of isotropic rectangular samples.

## IV. CONCLUSIONS

This work provides a simple method for measuring the electrical resistivity of rectangular isotropic samples. The simulation using COMSOL Multiphysics software allowed us to study the Montgomery method in much more detailed never reported before, especially with regard to the behavior of the electric potential and the electric field of the simulated samples. The simulations proved that the analytical method previously reported can be used in many practical situations. The results show that the method reported here can be easily used in substitution to the conventional four-probe method.

## ACKNOWLEDGMENTS

L. M. S. Alves is post-doc at EEL-USP (Proc. 21.1.419.88.8). F. S. Oliveira is post-doc at UNICAMP (FAPESP 2021/03298-7). M. S. da Luz is CNPq fellow (Proc. 311394/2021-3).

## AUTHOR DECLARATIONS (Conflict of Interest)

The authors have no conflicts to disclose.

## REFERENCES

[1] I. Miccoli, F. Edler, H. Pfnür, C. Tegenkamp, "The 100th anniversary of the four-point probe technique: the role of probe geometries in isotropic and anisotropic systems", Journal of Physics: Condensed Matter **27,** 223201 (2015).

[2] Y. Ju Yun, H. Young Yu, D. Han Ha, "Measurement of electrical transport along stretched l-DNA molecules using the four-probe method", Current Applied Physics 11, 1197 (2011).

[3] S. Matsuo, N. R. Sottos, "Single carbon fiber transverse electrical resistivity measurement via the van der Pauw method", Journal of Applied Physics **130**, 115105 (2021).

[4] F. Keywell and G. Dorosheski, "Measurement of the Sheet Resistivity of a Square Wafer with a Square Four-Point Probe", Review of Scientific Instruments **31**, 833 (1960).

[5] M. G. Buehler and W. R. Thurber, Measurement of the resistivity of a thin square sample with a square four-probe array, Solid-State Electronics 20, 40344 (1977).

[6] B. L. Theraja, R. S. Sedha, *Principles of Electronic Devices and Circuits* (S. Chand Publishing, 2007).

[7] S. M. Sze and K. K. Ng, *Physics of Semiconductor Devices* (Hoboken, NJ: Wile, 2007).

[8] F. Wenner, "A Method of Measuring Earth Resistivity", Bulletin of the Bureau of Standards 12, 469 (1915).

[9] L. B. Luganskya and V. I. Tsebro, "FourProbe Methods for Measuring the Resistivity of Samples in the Form of Rectangular Parallelepipeds", Instruments and Experimental Techniques 58, 118 (2015).

[10] I. Kazani, G. De Mey, C. Hertleer, J. Banaszczyk, A. Schwarz, G. Guxho, L. Van Langenhove, "van der Pauw method for measuring resistivities of anisotropic layers printed on textile substrates", Textile Research Journal. 81(20), 2117 (2011).

[11] W. Versnel, "Electrical characteristics of an anisotropic semiconductor sample of circular shape with finite contacts", Journal of Applied Physics. 54, 916 (1983).

[12] van der Pauw L J 1958 Philips Res. Rep. 13 1

[13] van der Pauw L J 1958 Philips Tech. Rev. 20 220

[14] H. C. Montgomery, "Method for Measuring Electrical Resistivity of Anisotropic Materials", Journal of Applied Physics 42, 2971 (1971).

[15] B. F. Logan, S. O. Rice, and R. F. Wick, "Series for computing current flow in a rectangular block", Journal of Applied Physics 42, 2975 (1971).

[16] J. D. Wasscher, "Note on four-point resistivity measurements on anisotropic conductors", Philips Research Reports 16, 301 (1961).

[17] C. A. M. dos Santos, A. de Campos, M. S. da Luz, B. D. White, J. J. Neumeier, B. S. de Lima, and C. Y. Shigue, "Procedure for measuring electrical resistivity of anisotropic materials: A revision of the Montgomery method", Journal of Applied Physics 110, 083703 (2011).

[18] F. S. Oliveira, R. B. Cipriano, F. T. da Silva, E. C. Romão, C. A. M. dos Santos, "Simple analytical method for determining electrical resistivity and sheet resistance using the van der Pauw procedure", Scientific Reports 10, 16379 (2020).

[19] S. H. N. Lim, D. R. McKenzie, M. M. M. Van der Bilek, "van der Pauw method for measuring resistivity of a plane sample with distant boundaries. Review of Scientific Instruments 80, 075109 (2009).

[20] ASTM 1975 Annual Book of ASTM Standards part 43, F84.

[21] R. A. Serway Principles of Physics (London: Saunders College Publishing, 1998).

[22] COMSOL™ 5.1 Multiphysics, Reference Manual, 2015.

[23] J. Lin, P. Y. Wong, P. Yang, Y. Y. Lau, W. Tang, P. Zhang, "Electric field distribution and current emission in a miniaturized geometrical diode", Journal of Applied Physics **121**, 244301 (2017.

[24] A. I. Sharshir, S. A. Fayek, A. F. Abd El-Gawad, "Experimental investigation of E-beam effect on the electric field distribution in cross-linked polyethylene/ZnO nanocomposites for medium-voltage cables simulated by COMSOL Multiphysics", Journal of Analytical Science & Technology **13**, 16 (2022).

[25] P. Song, Q. Song, Z. Yang, G. Zeng, H. Xu, X. Li, W. Xiong, "Numerical simulation and exploration of electrocoagulation process for arsenic and antimony removal: Electric field, flow field, and mass transfer studies", Journal of Environmental Management **228**, 336 (2018).

[26] J. Furlan, J. A. Martins, E. C. Romão, "Dispersion of toxic gases (CO and $CO_2$) by 2D numerical simulation", Ain Shams Engineering Journal **10**, 151(2019).

[27] C. A. Perussello, C. Kumar, F. de Castilhos, M.A. Karim, "Heat and mass transfer modeling of the osmo-convective drying of yacon roots (Smallanthus sonchifolius)", Applied Thermal Engineering **63**, 23 (2014).

[28] M. Daş, E. Alıç, E. K. Akpinar, "Numerical and experimental analysis of heat and mass transfer in the drying process of the solar drying system, Engineering Science and Technology, an International Journal **24**, 236 (2021).